\documentclass{aastex631}

\usepackage{mathtools}
\usepackage{multirow}
\usepackage{amsmath}

\graphicspath{{./}{figures/}}

\begin{document}

\title{High frequency decayless waves with significant energy in Solar Orbiter/EUI observations
}

\author[0000-0002-0175-7449]{Elena Petrova }
\affiliation{Centre for mathematical Plasma Astrophysics, Mathematics Department, KU Leuven, Celestijnenlaan 200B bus 2400, B-3001 Leuven, Belgium }

\author[0000-0001-5731-8173]{Norbert Magyar }
\affiliation{Centre for mathematical Plasma Astrophysics, Mathematics Department, KU Leuven, Celestijnenlaan 200B bus 2400, B-3001 Leuven, Belgium }

\author[0000-0001-9628-4113]{Tom Van Doorsselaere }
\affiliation{Centre for mathematical Plasma Astrophysics, Mathematics Department, KU Leuven, Celestijnenlaan 200B bus 2400, B-3001 Leuven, Belgium }

\author[0000-0003-4052-9462]{David Berghmans }
\affiliation{Royal Observatory of Belgium, Ringlaan -3- Av. Circulaire, 1180 Brussels, Belgium}

\begin{abstract}

High-frequency wave phenomena present a great deal of interest as one of the possible candidates to contribute to the energy input required to heat the corona as a part of the AC heating theory. However, the resolution of imaging instruments up until the Solar Orbiter have made it impossible to resolve the necessary time and spatial scales. The present paper reports on high-frequency transverse motions in a small loop located in a quiet Sun region of the corona.The oscillations were observed with the HRIEUV telescope (17.4 nm) of the EUI instrument onboard the Solar Orbiter. We detect two transverse oscillations in short loops with lengths of 4.5 Mm and 11 Mm. The shorter loop displays an oscillation with a 14 s period and the longer a 30 s period. Despite the high resolution, no definitive identification as propagating or standing waves is possible. The velocity amplitudes are found to be equal to 72 km/s and 125 km/s, respectively, for the shorter and longer loop. Based on that, we also estimated the values of the energy flux contained in the loops - the energy flux of the 14 s oscillation is 1.9 $kW \cdot m^{-2} $ and of the 30 s oscillation it is 6.5 $kW \cdot m^{-2}$. While these oscillations have been observed in the Quiet Sun, their energy fluxes are of the same order as the energy input required to heat the active solar corona. Numerical simulations were performed in order to reproduce the observed oscillations. The correspondence of the numerical results to the observations provide support to the energy content estimates for the observations. 
Such high energy densities have not yet been observed in decayless coronal
waves, and this is promising for coronal heating models based on wave
damping.
\end{abstract}


\section{Introduction} \label{sec:intro}

Ever since the realisation that the corona is much hotter than the solar surface (or the photosphere), the question has been how it is heated. In order to prevent the corona from cooling down in a matter of hours, it needs to be heated by $100-200 \:W \cdot m^{-2}$ for a Quiet Sun region and $ ~ 10^4 \: W \cdot m^{-2}$ for active regions \citep{Withbroe1977}. There are various proposed candidate mechanisms that could be responsible for the deposition of the indicated energy, particularly MHD waves \citep{Doorsselaere2020}. 

Decayless kink oscillations constitute one of the possible candidates for heating the corona. In general, kink oscillations are one of the MHD wave modes in the overdense plasma cylinder aligned with the magnetic field. It is also the only mode that shows the transverse displacement of the density structure that they exist in. 
The periods lie in a broad range of values from several seconds to hours, and this mode is detected in propagating and standing regimes. The latter can be classified in terms of decay - there are rapidly decaying oscillatory motions \citep{nakariakov1999,schrijver1999,aschwanden1999} and decayless oscillations \citep{nistico2013,Anfinogentov2013,tian2012,wang2012}. The decay is attributed to resonant absorption and the decay time ranges from two to four oscillation periods. 
These two regimes also differ in the displacement amplitude: decayless oscillations are usually characterized by low-amplitude with an amplitude value around the minor radius of the loop. In contrast, for the case of decaying oscillations, the amplitude reaches several minor radii.
Moreover, there is a difference in excitation mechanism: decaying oscillations are typically associated with coronal eruptions. However, this is not the case for the decayless oscillations that are observed throughout most of the active coronal regions. It was suggested that the oscillations are continuously excited by some driver. 
The omnipresence of the decayless oscillations makes them one of the potential mechanisms for coronal heating \citep{shi2021}. 

A statistical study of such decayless oscillations was performed by \cite{Anfinogentov2015}, who revealed the omnipresence of this phenomenon as well as the range of the periods of the observed oscillations. The range of the periods encompasses values from 1.5 to 10 minutes and an average displacement amplitude of 0.17 Mm. The fact that there is a strong correlation of oscillation period with the length of the loop in the observed cases supports the interpretation of the observed waves as standing kink waves. 
3D numerical simulations of the decayless kink oscillations driven with the footpoint driver in a straight, density-enhanced magnetic flux tube in low-$\beta$ plasma performed by \cite{Karampelas2019} suggest that the studied oscillations of small amplitude contain enough energy flux to support Quiet Sun regions. In this setup, the amplitudes of 0.15-0.55 Mm result in flux values of up to $110 \: W\cdot m^{-2}$. In the recent models of \citet{shi2021}, the decayless transverse waves provide sufficient energy to maintain a loop at coronal temperatures against radiative losses. 

However, the range of the observed oscillation periods is restricted by the capabilities of the imaging instruments, for SDO being a 12 s time resolution and 0.6 arcsec for the spatial. Therefore, detecting the oscillations with relatively short periods is not possible for imaging instruments with such cadence.  

Higher frequency oscillations were previously accessible only via radio observations or eclipse data. For example, \cite{Williams2002} detected a 6-s oscillation in an active region coronal loop. The temporal evolution of the intensity was one of the factors supporting the interpretation of the observed wave as propagating fast magneto-acoustic mode. Later \cite{Samanta2016} detected high-frequency oscillations with periods ranging from 6 to 25 seconds from multi-slit spectroscopic observations of the corona during the eclipse as well. The oscillations are detected in signals of intensity, Doppler velocity, and line width. The oscillations of the Doppler velocity alone are attributed to the fast kink mode, while the oscillations of the intensity and velocity are interpreted as compression waves.
It follows that the eclipse data is convenient to study high-frequency wave properties of the corona and outperform the majority of space-based telescopes. However, on top
of the inherent sparse eclipse observations and short time duration, these eclipse instruments lack spatial resolution to identify wave modes with great certainty.
In the radio, fast coronal oscillations with a period of 3s were previously detected \citep{Kolotkov2018}. Unfortunately, radio observations suffer the same lack of spatial resolution as eclipse observations.

Knowing the energy contained in a different range of frequencies would help to build the power law as it was made by \cite{Morton2016} and \cite{Morton2019} for the frequencies in the low range - 0.2 to 11 mHz with the use of the CoMP data. The power was calculated for different magnetic structures - active regions, open field lines, and quiet sun for the specified range of frequencies. The distribution of power depending on the frequencies and regions was found to be the following: for the lower frequencies, the power is more significant in the active regions, while for the higher frequencies, the greater power is contained in the quiet sun region compared to the active regions. 

 
The interpretation of the observed transverse waves was subject to a debate \citep{vd2008} - whether to interpret them as bulk Alfvén waves or kink waves (sometimes referred to as surface Alfvén waves). The latter received stronger support, and the consensus is reached - transverse displacement of the magnetic cylinder driven by tension forces with the phase velocities belonging to the interval between the internal and external Alfvén speeds can be interpreted as surface Alfvén waves \citep{Goossens2012}. This has consequences on the energy calculation since the energy is localized in the structure for the kink waves, while in bulk Alfvén waves, the energy has a uniform distribution in space. Non-uniformity affects the choice of the values that one operates with - averaged energy values are used instead of the energy density. 
For the case of the single flux tube, misinterpretation leads to the overestimation of the energy in the observed waves and needs to be corrected \citep{Goossens2013}. For a bundle of tubes, the so-called filling factor needs to be taken into account as it was introduced by \cite{Doorsselaere2014}.

The oscillatory phenomena for which the associated energy fluxes are estimated are found in coronal regions: Quiet Sun region and coronal holes observed by SDO with values of wave energy flux of 
$ 100-200 \: W m^{-2}$   \citep{McIntosh2011}, active region coronal loops observed by CoMP \citep{Tomczyk2007} with energies $ 0.01 \: W m^{-2}$, transverse waves in solar plumes observed by SDO with energies constituting $ 9 - 24 \: W m^{-2}  $ \citep{Thurgood2014}; chromosphere - transverse oscillations spicules
\citep{DePontieu2007} with energies $4-7 \:  kW m^{-2}$; photosphere - oscillations associated with a large bright-point group with periods in the range of 126 - 700 seconds and energy flux in the chromosphere of $15000 \: W m^{-2} $ \citep{Jess2009}.
The values given above for the energy flux in coronal regions are considerably lower than the values needed to compensate for the coronal energy losses \citep{Withbroe1977}. Therefore, there should be other ways to supply additional energy. 

The current paper addresses the new observations made by the Solar Orbiter and presents the results of the analysis of detected high-frequency transverse oscillations seen in the 17.4 nm channel. The data used for the analysis is obtained from one of the three telescopes of the Extreme Ultraviolet Imager (EUI) \citep{Rochus2020} onboard the Solar Orbiter satellite \citep{Marirrodriga2021} launched in February 2020. Its trajectory is designed in a way that allows imaging active regions for a significant amount of time with a perihelion below 0.3 AU, and an inclination enough to observe the polar regions and at the same time minimize the distance to the Earth to make sure that data return is optimized .

EUI includes the High Resolution Imager (HRIEUV), which observes in the EUV, at 17.4nm. Apart from the latter, the \mbox{17.4 nm}, the EUI is accompanied by the channel corresponding to the Lyman-$\alpha$ line. During the perihelion, the imagers with the one arcsec resolution provide imaging observations where 1 pixel corresponds to $(100 \mbox{km})^2$ on the Sun. The temporal capability of both instruments reaches an imaging rate of 1 Hz.

The combination of such temporal and spatial resolution is quite remarkable as up to present, the data with similar sampling characteristics was only obtained for a short period of time via rockets Hi-C \citep{Kobayashi2014} and VAULT \citep{Korendyke2001}.
Furthermore, high cadence allows access to high-frequency phenomena and gives a promising opportunity to evaluate the energy content. 

The details of the observations are given in the Section~\ref{sec:obs}, estimation of the energy content in Section~\ref{sec:energy}. Section~\ref{sec:simluations} is devoted to the comparison of the numerical simulations with observations. In Section~\ref{sec:discussion} we summarize and discuss the results of the findings.

\section{Observations and data analysis} \label{sec:obs}

\begin{figure}[ht!]
\epsscale{0.85}
\begin{interactive}{animation}{feb23_movie_new.mp4}
\plotone{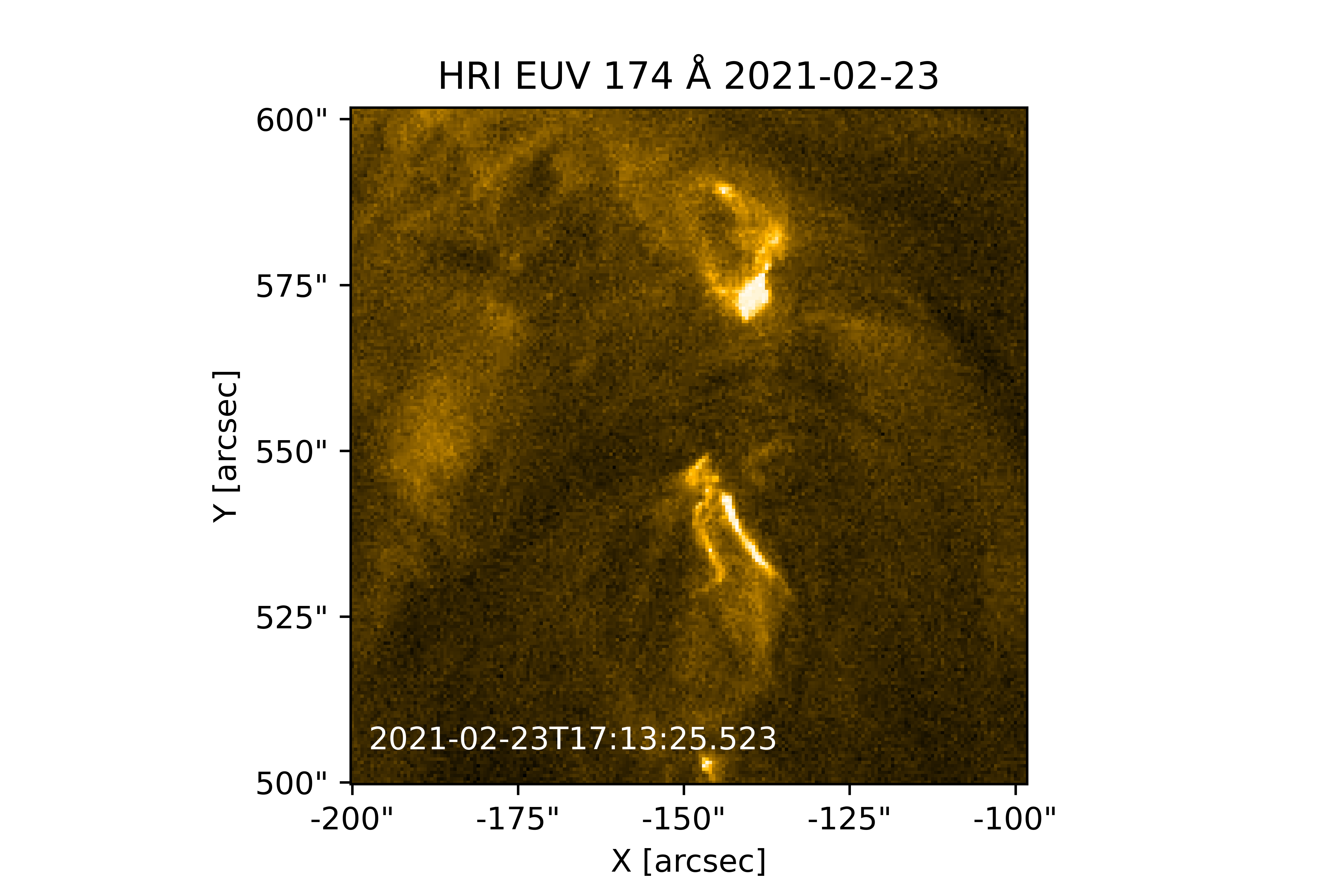}
\end{interactive}
\caption{ Sequence of images taken by HRIEUV. It covers 7.5 minutes of observing beginning 
at 17:13:25 UT on 2021 February 23.  
\label{fig:video1}}
\end{figure}

The images were obtained on 2021 February 23 from 17:13:25 to 17:20:59, with a cadence of 2 seconds\texttt{}\footnote{https://doi.org/10.24414/k1xz-ae04}. At that time, SolO was located
at a distance of 0.52 AU from the Sun. In this setting, a pixel corresponds to $200\times 200 \: km^2$ on
the Sun. Moreover, Solar Orbiter was more than 150 degrees in solar longitude separated from
Earth, and as a result, there are no simultaneous AIA observations to compare with.
As a preparation step for the analysis, the images taken at different times were co-aligned using \texttt{SunPy}\footnote{https://sunpy.org} \citep{sunpy2020} embedded function \textit{ mapsequence\textunderscore coalign\textunderscore by\textunderscore match\textunderscore template} which takes as argument shifts calculated relative to the first map. These shifts are calculated with cross correlation in two iterative steps. Also independent IDL-based routines were exercised to confirm the co-alignment. The apparent stability of the resulting movie (Fig~\ref{fig:video1}) validates that the influence of telescope jitter on the results will be minimal.

The sequence of the images as an animation is shown in Fig~\ref{fig:video1}. The region of interest lies in $[-150'',  -125 '']$ on the x-axis and $[525'', 550'']$ on the y-axis. The coordinates of field of view are given in  helioprojective-cartesian coordinates as seen from Solar Orbiter.
The configuration of the region with detected oscillations is shown in Fig~\ref{fig:fig2}. 
There are two loops analyzed - loop A and loop B 
represented by red dashed lines in figures Fig. \ref{fig:fig2} (a) and (e). They are visible on the disk and rather well-contrasted with respect to the background. The loop system has a very dynamic behavior. Loop A is formed in a sigmoid shape and is ejected away to the left side after some evolution, therefore it is no longer visible after 80 seconds. Loop B is detectable only after loop A's disappearance as loop B seems to be underlying.

\begin{figure}[h!]
\gridline{\fig{EUI_loopA_t0}{0.3\textwidth}{(a)} \fig{loopA_t0}{0.25\textwidth}{(b)}
\fig{loopA_t14}{0.25\textwidth}{(c)}
\fig{loopA_t28}{0.25\textwidth}{(d)}
}

\gridline{\fig{EUI_loopB_t90}{0.3\textwidth}{(e)} \fig{loopB_t90}{0.25\textwidth}{(f)}
\fig{loopB_t100}{0.25\textwidth}{(g)}
\fig{loopB_t110}{0.25\textwidth}{(h)}
}

\caption{The 17.4 nm image taken by HRIEUV on 23 February 2021. The two oscillating loops are highlighted with red dashed lines and shown by figures (a) - loop A and (e) - loop B. The white lines show the location of the slits. The blue rectangle shows the region for the zoomed view. The (b), (c) and (d) figures show the zoomed view of the loop A for the time steps of 0, 14 and 28 seconds.  Panels (f), (g) and (h) show the zoomed view of the loop B  for the time steps of 90, 100 and 110 seconds. 
}
\label{fig:fig2}

\end{figure}

\begin{figure}[h!]
\epsscale{0.65}
\begin{interactive}{animation}{feb23_movie_sub_cut.mp4}
\plotone{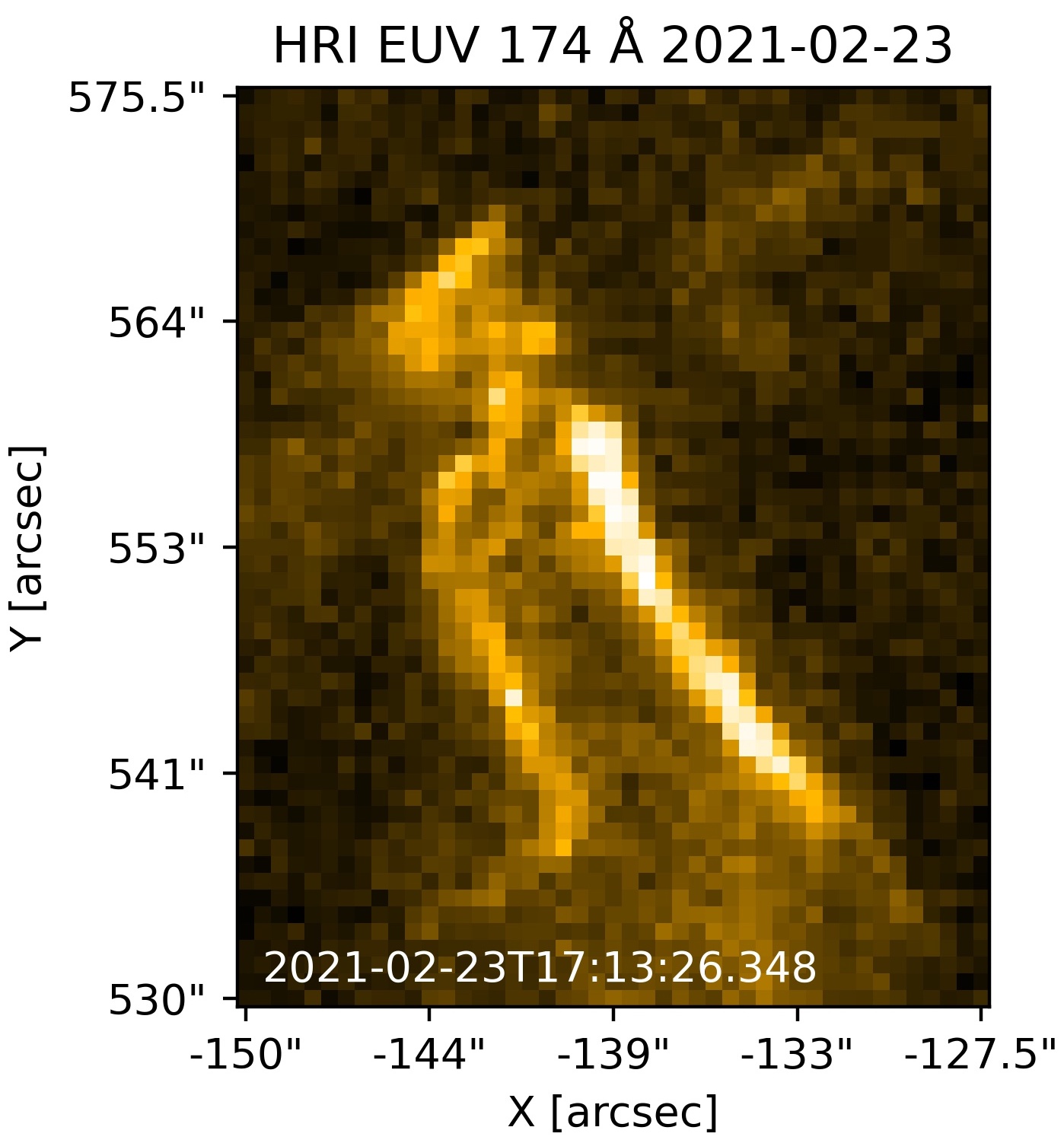}
\end{interactive}
\caption{ Sequence of images taken by HRIEUV with the zoomed-in field of view covering 140 seconds from the first frame, where the oscillations are present. 
\label{fig:video_zoom}}
\end{figure}

The movie with the zoomed-in field of view is shown in Fig. \ref{fig:video_zoom}.

The shape of the loop is assumed to be semi-circular, so the length of the loop is calculated as $ L=\pi r $. The $r$ value is calculated using two different methods: the first - the form of the loop is fitted by the manually clicked points, and then the distance between the consecutive points was calculated and summed up. These lengths are reported in Table \ref{tab:table1}. The second method - the location of the two footpoints was identified by visual inspection and then $2r$ calculated as the distance between them. Obviously, the second method results in a lower value which is indicated in brackets in the second row of Table \ref{tab:table1}. Two different values of the length of the loop result in two different values for all the parameters, including the length as a variable. Therefore, there are two different values of the phase speed where the value in brackets corresponds to the second method of the loop length calculation. The value of the phase speed is in turn calculated assuming that what is observed is a fundamental mode of a standing kink wave. Accordingly, the wavelength is determined by the double loop length. 

To more quantitatively track the loop oscillations we fitted a Gaussian profile to the intensity at each time frame in the slits indicated by the white line in Fig.~\ref{fig:fig2}. Additionally, the parameters of the Gaussian fitting, particularly the $\sigma$ value, are used to calculate the radius of the cross-section of the loop. The parameter  $\sigma$ and the half width at half maximum (HWHM) are connected via a linear relation. Therefore, the radius of the loop is calculated as HWHM and is determined by: $\mbox{HWHM} = \sqrt{2 \ln{2}}\sigma$. 
Further, the amplitude and period of the oscillations were identified based on the fitting of the sine function to the loop central position at each point of time: 
\begin{equation}
    x(t) = A \sin(\frac{2\pi t}{P}+\phi) + h
\end{equation}
 Here the transverse displacement amplitude is defined by $A$, $P$ is the period of oscillation, $\phi$ is the phase, and $h$ is a constant value. The velocity amplitude can be calculated for a given displacement and period as $v = 2 \pi A P^{-1}$.

The main parameters of the loops, as well as characteristics of the wave phenomena detected, are shown in the Table~\ref{tab:table1}. 

\begin{table}[h!]
\centering
\caption{Key parameters of the oscillating loops }
\begin{tabular}{|l|c|c|}
\hline
Characteristics             & \multicolumn{1}{l|}{30 s oscillations (loop A)} & \multicolumn{1}{l|}{Super short oscillations (loop B)} \\ \hline
Period $P$, s                  & $30\pm 0.5$         & $14\pm 0.15 $        \\ \hline
Length of the loop $L$, Mm        & $11.7\pm 1.2$ ($9.3 \pm 1$) & $4.53\pm 0.5$ ($4.05\pm 0.4$) \\ \hline
Radius of the loop cross-section $R$, Mm & $0.38\pm0.03$ & $0.327\pm0.03$ \\ \hline
Displacement $A$, km             & $600\pm 50$        & $160\pm30$       \\ \hline
Velocity amplitude $v$, km/s & $125.6\pm10.7$      & $71.8\pm11$        \\ \hline
Phase speed $V_{ph}$, km/s & $780.5\pm 80$ ($620.2 \pm 63$)                         & $647.3\pm65$ ($579.2\pm58$)                                 \\ \hline
Energy flux, $W\cdot m^{-2}$ & $6529.6\pm1448$      & $1990.7\pm680$\\ \hline
Energy density, $J\cdot m^{-3}$ &  $0.0105\pm0.002$      & $0.0034\pm0.001$      \\ \hline

\end{tabular}
\label{tab:table1}
\end{table}

The fitted positions of the centroids together with the fitted sine function are shown in Fig.~\ref{fig:fig3_30} and Fig.~\ref{fig:fig3} which represent time distance maps along horizontal slits crossing the loops A and B. 

\begin{figure}[h!]
\gridline{\fig{30s_slit0}{0.6\textwidth}{(a) }\fig{Er30s_0}{0.2\textwidth}{(b)}}
\gridline{\fig{30s_slit1}{0.6\textwidth}{(c)}\fig{Er30s_1}{0.2\textwidth}{(d)}}
\gridline{\fig{30s_slit2}{0.6\textwidth}{(e)}\fig{Er30s_2}{0.2\textwidth}{(f)}}
\gridline{\fig{30s_slit3}{0.6\textwidth}{(g)}\fig{Er30s_3}{0.2\textwidth}{(h)}}
\caption{Time distance maps of the oscillating loop A. The fitted profile positions are indicated by the red points. The dashed blue curve represents the fitted sinusoidal function. The highest panel of the left column corresponds to the northernmost slit, and the lowest panel figure (g) correspond to the southern slit. The right column corresponds to zoomed view with the error bars. The region for the zoomed view is indicated by the blue rectangle in figure (a).}
\label{fig:fig3_30}

\end{figure}

\begin{figure}[h!]
\gridline{\fig{14s_slit0}{0.6\textwidth}{(a) }\fig{Er14s_0}{0.2\textwidth}{(b)}}
\gridline{\fig{14s_slit1}{0.6\textwidth}{(c)}\fig{Er14s_1}{0.2\textwidth}{(d)}}
\gridline{\fig{14s_slit2}{0.6\textwidth}{(e)}\fig{Er14s_2}{0.2\textwidth}{(f)}}
\gridline{\fig{14s_slit3}{0.6\textwidth}{(g)}\fig{Er14s_3}{0.2\textwidth}{(h)}}
\caption{Time distance maps of the oscillating loop B. The fitted profile positions are indicated by the red points. The dashed blue curve represents the fitted sinusoidal function. The highest panel of the left column corresponds to the northernmost slit, and the lowest panel figure (g) correspond to the southern slit. The right column corresponds to zoomed view with the error bars. The region for the zoomed view is indicated by the blue rectangle in figure (a).}
\label{fig:fig3}

\end{figure}

As can be noted from Fig.~\ref{fig:fig3_30} and Fig.~\ref{fig:fig3}, the oscillations are not present in simultaneous time frames. The oscillations in the loop A are present from the first time frame and last for 80 seconds, while the oscillations in loop B are only detected after 90 seconds from the start of the imaging. 
The sequence of the images depicts the configuration of the loops more clearly, showing that for the loop displaying 30 s oscillations, it is not possible to detect more than two periods of the oscillations, before the loop is ejected away to the left side after 80 seconds.
From Fig.~\ref{fig:fig3} it could have been inferred that one loop shows both transverse oscillations. However, tracking the fitted points on the sequence of the images clearly indicates the presence of two different loops, with the second only becoming visible after the disappearance of the first one. 

Both oscillations are tracked in several neighboring slits, while for the longer period, the width of the slit is 3 pixels in order to increase the signal, while for the 14 s it is only 1 pixel. The latter setting for the shorter period oscillation is chosen due to the limited amount of slits where the oscillations are detected.  

In order to investigate whether the observed waves are propagating or standing, the fitted sine positions were followed from slit to slit for both of the detected oscillations. The consistent slope of the location of the peaks relative to the previous ones would indicate the presence of the propagating wave. However, no evidence of propagating waves was detected for both of the analyzed oscillations since there is no notable shift of the followed consecutive peaks. However, given the cadence of the data images and the distance between the slits, it would be hard to distinguish propagating waves from standing. For example, considering the 14 s oscillations, the slit width is 1 pixel, corresponding to 200 km in physical units. There are four slits where the oscillations are detected. In the case of a propagating wave, the peak with the phase speed of $650\: \mbox{km/s} $ will move from one slit to the consecutive in 0.3 seconds, while the cadence is 2 seconds, so it would not be possible to make a definitive conclusion. 

The observed waves are identified as decayless kink oscillations because the time-distance diagrams show loops oscillating transversally with low amplitudes with no apparent decay. The decay time of the decaying oscillations ranges from two to four periods. Since the oscillations of loop A are observed for two periods in several slits and the oscillations of loop B are observed at least for three periods, and no decay was identified in those fits, we can conclude that the nature of the observed waves is decayless. 
The period and wavelength match the linear scaling that was derived with SDO/AIA \citep{Anfinogentov2015}. 

The calculated values of the periods and amplitudes are influenced by measurements errors. The starting point of calculating error bars is the estimation of the uncertainty in the data itself. The estimated for HRIEUV high gain is 
\begin{equation}
    \sigma^2 = (read\:  noise)^2 +(Intensity) \cdot 6.85
\end{equation}

Here read noise is estimated to be equal to 2 DN (digital numbers), intensity is also given in the same units from the Level 1 data. 
Further, these estimations are used as an uncertainty in data for the gaussian fitting. The error bars representing the uncertainty, are shown in Fig.~\ref{fig:fig3_30} and Fig.~\ref{fig:fig3}. The values of the propagated errors are given in Table \ref{tab:table1}.

\section{Energy content} \label{sec:energy}

As was mentioned above, the interpretation of the observed transverse waves as kink waves better fits the observed phenomena \citep{Goossens2013}. Therefore, the total energy (in $J$) in the kink wave, which is related to the total energy of bulk Alfvén wave, has to be calculated according to \citep{Doorsselaere2014}:

\begin{equation}
TE = \frac{1}{2}V_{loop}v^2(\rho_i+\rho_e)
\label{energy}
\end{equation}

Here the $V_{loop}$ is the volume of the loop is calculated as $V_{loop} = \pi R^2 L$. The $\rho_i$ and  $\rho_e$ are the densities inside and outside the loop, respectively. The density inside the loop is for the quiet solar corona estimated as an order of magnitude $10\cdot10^{-13}\: kg \cdot m^{-3}$. The density contrast is assumed to be equal to 3, therefore, the value of the external density $\rho_e$ is of the order of  $3.3\cdot10^{-13}\: kg \cdot m^{-3}$. The total energy over the volume of the loop will give the energy density (in $J m^{-3}$). Finally, the energy flux is the product of the energy density and group speed at which the energy is propagated. For the kink oscillations, the phase speed and the group speed can be considered as equal. The energy flux of the 14 s oscillation is 1991 $W \cdot m^{-2} $ and of 30 s is 6530 $W \cdot m^{-2}$. The results are given in the last row of Table~\ref{tab:table1}. 

It must be noted that the orders of the calculated values are corresponding to the order of the energies required to heat the corona, specifically for the 30 s period oscillations.
Even with the observational uncertainties and density guess, these measurements have the correct order of magnitude to compensate for the radiative losses \citep[as put forward by][]{Withbroe1977}. Certainly their energy content is at least 1-2 orders of magnitude larger than any previous estimates for coronal wave energy content.


Kink oscillations are shown to be prone to the development of the Kelvin-Helmholtz instability in the loop's boundary \citep{Terradas2008}. KHI results in the evolution of turbulence, which is efficient in forming small scales to dissipate the wave energy and therefore heat coronal loops \citep{karampelas2017,shi2021}. The method to model decayless kink oscillations to recover the observational parameters is as energy injected in a system through the footpoints. Assuming steady-state turbulence - all the energy injected cascades and dissipated at the same rate it is injected. Therefore, for the specific value of the energy injected into the system, the value of oscillation amplitude required to dissipate all the energy by the turbulence, was calculated by \citet{Hillier2020}. In this setting, the injected energy flux scales as a third power of the velocity amplitude. With this mentioned, the injected energy flux towards the small scales can be estimated from the velocity amplitudes, following the route described above, but in the opposite direction. For the velocity amplitudes in the current analysis, the values of the energy cascade rates are $ 98.16 \: W \cdot m^{-2}$ and $9.25 \: W \cdot m^{-2}$ for the lower and higher frequency oscillation, respectively. Contrary to the total energy content, these values are not comparable to the energy flux values calculated for the current loops and indicated in the last row of the table since it is a value of the input energy injected into the system such that it is fully dissipated by the turbulence. However, if the energy is injected with the observed rate of $6.5 kW \cdot m^{-2}$ and $1.9 kW\cdot m^{-2}$, respectively, then this energy supply acts as a reservoir to sustain the turbulent cascade for a longer time. Note that the energy cascade operates on a much shorter time than resistive processes \citep{arregui2015,terradas2018b}.

The calculated energy budget can be compared to the radiative losses, which is one of the primary energy loss mechanisms in the corona. 
In order to estimate what portion of the energy contained in the loop is radiated, we can model the radiative losses as $\chi \rho^2  T^{\beta}$ for which the temperature dependence and specific values of the parameters are determined from the CHIANTI database. 
For a given temperature and coronal plasma concentration, the value of the radiative losses approximately equals  $2\cdot 10^{-4} \: W \cdot m^{-3}$ which can be compared to the energy density of 30 s oscillating loop that constitutes  $  3.509\cdot 10^{-4} \:  W \cdot m^{-3}$ and for the 14 s oscillating loop - $2.45  \cdot 10^{-4} \: W\cdot m^{-3}$. This fact indicates that a substantial amount of energy is available on top of the radiative energy losses that can heat the plasma.

\section{Numerical simulations}
\label{sec:simluations}

In this subsection we aim to reproduce the high frequency kink oscillations of loop A through numerical simulations. On the one hand, if similar oscillation parameters are obtained through simulations, it reassures that the observed oscillations are indeed kink modes. Specifically, we model them as the fundamental standing kink mode. On the other hand, parameters determined through the observations, such as the energy content, can be verified, and additionally new parameters, such as the average magnetic field, can be estimated. \par 
We solve the full 3D ideal MHD equations using \texttt{MPI-AMRVAC}\footnote{http://amrvac.org} \citep{2018ApJS..234...30X,2020arXiv200403275K}. The $(x,y,z)$ extents of the numerical box are $(-20,20)  \times (-15,15) \times (0,30) \mathrm{Mm}$, with $z = 0$ being the photospheric height. We prescribe a realistic solar atmosphere by numerically integrating the hydrostatic equation using the FAL temperature profile $T(z)$ from \citet{2007ApJ...667.1243F}:
\begin{equation}
    \frac{d p(z)}{d z} = - g \rho (z), \rho(z) = \frac{m\ p(z)}{k_B T(z)},
    \label{hydrostatic}
\end{equation}
where $g = -274\ \mathrm{m\ s^{-2}}$ is the gravitational acceleration at the surface of the Sun, $m \approx 0.62\ m_p$ is the mean mass per particle for photospheric abundances with $m_p$ the proton's mass, and $k_B$ is Boltzmann's constant. In Equation~\ref{hydrostatic} an ideal equation of state is assumed. \par 
As described previously, the observed loops display a sigmoid appearance. Kink oscillations in sigmoid coronal loops were modeled previously by \citet{2020ApJ...894L..23M}. Here we adopt their magnetic field configuration and loop construction technique. Therefore, using their Eqs.(1)-(3) for a dipolar and helical magnetic field configuration, we set the parameters $z_0 = 2\ \mathrm{Mm}, d = 1.5\ \mathrm{Mm}$ and $\alpha = 1$. The coronal loop is added by tracing a single magnetic field line originating at $(x,y,z=0) = (2.5,-0.5)\ \mathrm{Mm}$, with a density ratio of three between the loop and the external plasma, at the base of the corona. The base numerical resolution is $128 \times 96 \times 96$. We use four levels of refinement, with the refinement criteria being the density in Löhner’s error estimator, and only applied to the coronal part of the simulation. The impulsive velocity perturbation at $t=0$ is defined in the same volume as the loop, varying sinusoidally along it in order to perturb a fundamental kink mode. The velocity vector is  horizontal and perpendicular to the loop axis in the apex, with this direction not changing along the loop. We set zero-divergence, open boundary conditions at all boundaries. \par
\begin{figure}[h!]
\begin{interactive}{animation}{FoMo_movie.mp4}
\epsscale{0.8}
\plotone{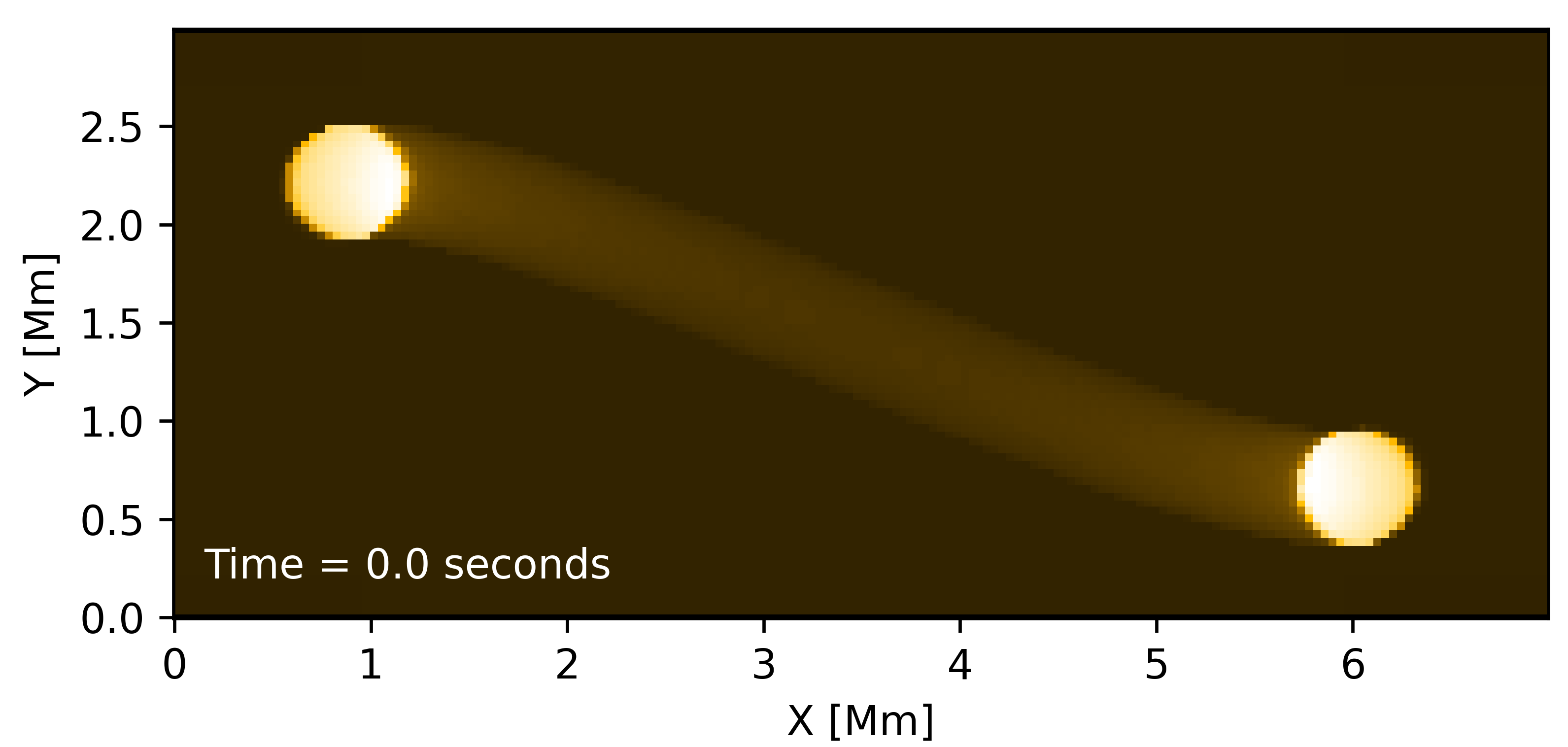}
\end{interactive}
\caption{ Sequence of forward-modeled 17.1 nm intensity images. 
\label{fig:video2}}
\end{figure}

 
In order to analyze the simulations in the same fashion like the observations, we output forward-modeled images of the loop oscillation in 17.1 nm, integrated along the $z$-axis of the simulation box and with a resolution of 10 km/pixel. The sequence of forward-modeled images is shown in Fig \ref{fig:video2}.

\begin{figure}[h!]
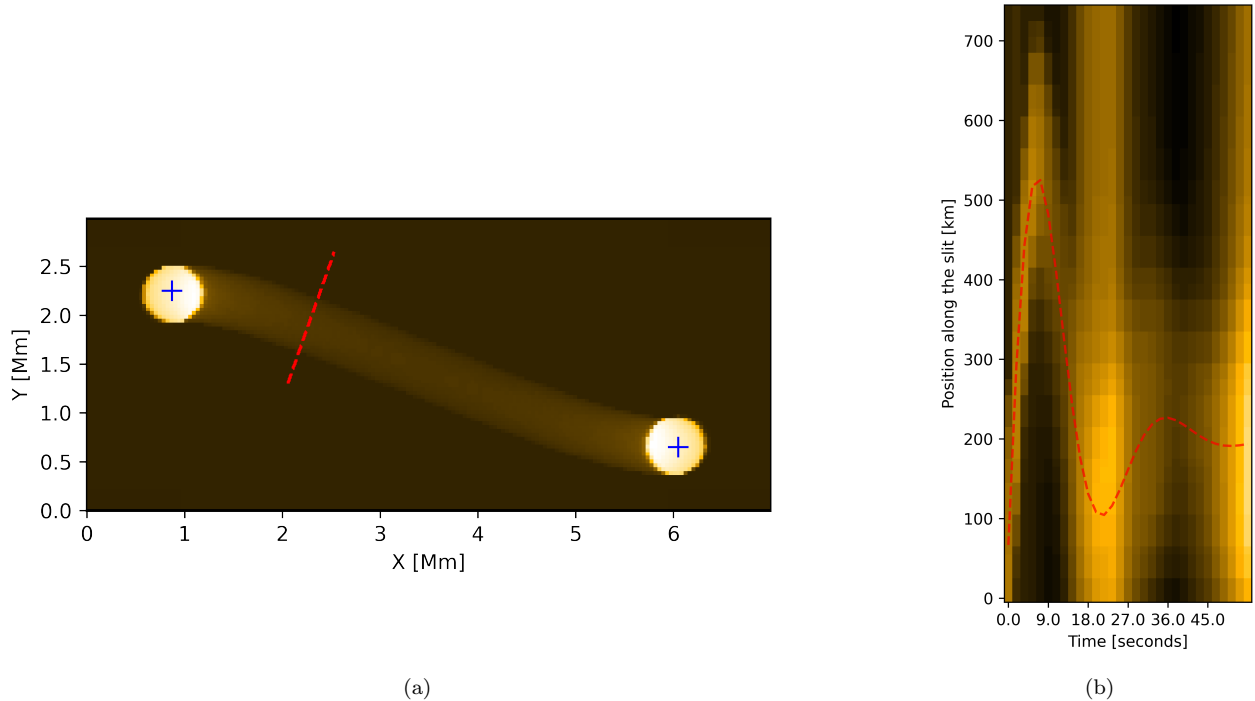

\gridline{\fig{FoMo_slit_location_171}{0.65\textwidth}{(a)} \fig{FoMo_oscillations_171}{0.25\textwidth}{(b)}}
\caption{(a) Forward-modeled intensity image of the initial configuration of the loop. The red dashed line shows the location of the slit. Blue crosses show identified locations of the footpoints of the loops. (b) The time-distance map of the slit with fitted sin function shown as a red dashed line. }
\label{fig:fig5}
\end{figure}

Loop oscillations in simulations are intended to mimic the oscillations of loop A from EUI observations. 
As seen from the sequence, the perturbation results in a periodic displacement of the loop with oscillations that damp rather quickly due to only one pulse put to the driver, while the observed waves are suggested to be driven continuously. 
In order to compare the simulation results, the forward-modeled images were processed in the same way as observational images. The slit location is chosen perpendicular to the loop axis, where the oscillations are more prominent. The slit location is shown in Fig \ref{fig:fig5} (a). The corresponding time-distance map with the fitted position of centroids is shown in Fig \ref{fig:fig5} (b). Locations for the loop center positions were found in the same way as for observations - first, a Gaussian was fitted to the intensity for each time step, and then the obtained positions were fitted using the sine function. However, contrary to the case of observations,  the oscillations damp very fast. Therefore, the sin function is multiplied by the factor of exponential decay.

The loop length is calculated using the second method in the analysis of the observations as the locations of the loop footpoints are clearly visible and were identified using visual inspection. As before, the shape of the loop is assumed to be semicircular. 
The main characteristics of the oscillations present in the forward modelling images are given in Table \ref{tab:table2}.

\begin{table}
\centering
\caption{Comparison of simulation results with observational characteristics}
\label{tab:table2}
\begin{tabular}{|l|c|c|c|}
\hline
Characteristics                          & SolO loop A       & forward modelling & input parameter   \\ \hline
Period $P$, s                            & 30         & 29  &    \\ \hline
Length of the loop $L$, Mm               & 11.7 (9.3) & 8.5 & 7.8    \\ \hline
Radius of the loop cross-section $R$, Mm & 0.38       & 0.3  & 0.3  \\ \hline
Displacement $A$, km                     & 600        & 618.9 &  \\ \hline
Velocity amplitude $v$, km/s             & 125.6      & 133.87 & 247 \\ \hline
Energy flux, $W \cdot m^{-2}$              & 6530       & 7005 & \\ \hline
Energy density, $J \cdot m^{-3}$              & 0.0105       & 0.0119 & 0.032\\ \hline

\end{tabular}
\end{table}

The value of the energy flux is calculated according to Eq.~\ref{energy} taking input values estimated from the forward modeled images with the same density assumptions. 
In the simulations, the velocity amplitude is slightly bigger due to displacement, and that results in a slightly higher value of the energy flux. 
In the table, the initial energy that was excited in the loop is given as $0.032 J m^{-3}$ in the initial velocity kick.
It is computed for the peak value at t=0, assuming the equipartition of the energy. The energy density value detected in the forward modeling is almost a factor of 3 higher than the input. A large fraction (around 50\%) of this initial energy leaves the system through leaky waves. The energy estimated from forward modelling thus coincides well with the remaining energy in the system after the initial phase. 
Similarity of the energy flux values supports the energy estimations from the observations, confirming that the observed high-frequency waves have high energy content in them, when compared to radiative losses.

\section{Discussion and conclusion} \label{sec:discussion}

The current study reveals that there are decayless transverse oscillations present with periods outside of the range of the previously detected decayless kink oscillations \citep{Anfinogentov2015}. The oscillations are detected in two short loops with lengths $11\pm1.2\: \mbox{Mm}$ and $4.5\pm0.5 \: \mbox{Mm}$ with velocity amplitudes of $125\pm10.7 \: \mbox{km/s}$ and $72\pm11\: \mbox{km/s}$ for the longer and shorter period oscillations, respectively. 
No definitive conclusions regarding the matter if it is a propagating or standing wave observed can be drawn from the current observations. 
Apart from the short periods, an important finding is the wave energy contained in the loop, whose values are the same order as the amount of energy required to heat the Quiet sun regions or even an active region loop. The energy flux of the 14 s oscillation is $1.9\pm0.68\: kW \cdot m^{-2} $ and of 30 s is $6.5\pm1.4\: kW \cdot m^{-2}$. 

Despite the fact that the estimates of the energies are suffering from an uncertainty originating in the observational side and resulting in approximate estimates of the loop and oscillation parameters, the energy budget contained in the detected oscillations is sufficient to compensate for the coronal radiative losses and constitute one of the possible candidates to explain coronal heating. It can be concluded that the energy budget found in the analyzed oscillations is higher than what was previously found. Our results suggest that there is potentially a lot of energy at small scales and short periods.
Moreover, reproduction of the energy flux values in numerical simulations supports the observational energy estimates, providing strong evidence that high energy content is present in the detected high-frequency oscillations. 

Decayless ways were previously shown to be omnipresent throughout the active coronal regions, however, for the detected waves, the question of commonality remains yet unsolved. 
On the one hand, the interpretation as decayless kink waves would suggest that these high-frequency, small-scale waves occur everywhere in the solar corona, with sufficient power. If confirmed, they will provide an excellent candidate to play a key role in coronal heating. Moreover, it would mean that the power-law behavior of wave energy \citep{Morton2016} is severely influenced by the limited spatial resolutions and would need to be scaled up by orders of magnitude. 
On the other hand, on the observations from the Solar Orbiter that are available so far, similar events were not detected. Certainly, it does not necessarily mean that the detected event is unique in its nature. However, it can be that in the current situation, the observational conditions were somewhat exclusive, and they might not always be as favorable.
Earlier high-frequency phenomena with similar time resolution were made only using the radio and eclipse data. However, this is the first time high-frequency oscillations are detected from space with a spatial resolution that drastically outperforms the previous instrument's capabilities used for eclipse observations. 
Also, among the other questions to be answered is how these high-frequency waves would manifest themselves in spectroscopic data. 
Further analysis of Solar Orbiter observations will help to examine the omnipresence of the detected waves and yield critically
a better grasp on the available wave power in this high-frequency range.

\begin{acknowledgments}

Solar Orbiter is a space mission of international collaboration between ESA and NASA, operated by ESA. The EUI instrument was built by CSL, IAS, MPS, MSSL/UCL, PMOD/WRC, ROB, LCF/IO with funding from the Belgian Federal Science Policy Office (BELSPO/PRODEX PEA 4000112292); the Centre National d’Etudes Spatiales (CNES); the UK Space Agency (UKSA); the Bundesministerium für Wirtschaft und Energie (BMWi) through the Deutsches Zentrum für Luft- und Raumfahrt (DLR); and the Swiss Space Office (SSO).

TVD was supported by the European Research Council (ERC) under the European Union's Horizon 2020 research and innovation programme (grant agreement No 724326) and the C1 grant TRACEspace of Internal Funds KU Leuven. EP has benefited from the funding of the FWO Vlaanderen through a Senior Research Project (G088021N).

\end{acknowledgments}


\newpage
\bibliography{References,refs_tomvd}{}

\begin{thebibliography}{}
\expandafter\ifx\csname natexlab\endcsname\relax\def\natexlab#1{#1}\fi
\providecommand{\url}[1]{\href{#1}{#1}}
\providecommand{\dodoi}[1]{doi:~\href{http://doi.org/#1}{\nolinkurl{#1}}}
\providecommand{\doeprint}[1]{\href{http://ascl.net/#1}{\nolinkurl{http://ascl.net/#1}}}
\providecommand{\doarXiv}[1]{\href{https://arxiv.org/abs/#1}{\nolinkurl{https://arxiv.org/abs/#1}}}

\bibitem[{{Anfinogentov} {et~al.}(2013){Anfinogentov}, {Nistic{\`o}}, \&
  {Nakariakov}}]{Anfinogentov2013}
{Anfinogentov}, S., {Nistic{\`o}}, G., \& {Nakariakov}, V.~M. 2013, \aap, 560,
  A107, \dodoi{10.1051/0004-6361/201322094}

\bibitem[{{Anfinogentov} {et~al.}(2015){Anfinogentov}, {Nakariakov}, \&
  {Nistic{\`o}}}]{Anfinogentov2015}
{Anfinogentov}, S.~A., {Nakariakov}, V.~M., \& {Nistic{\`o}}, G. 2015, \aap,
  583, A136, \dodoi{10.1051/0004-6361/201526195}

\bibitem[{{Arregui}(2015)}]{arregui2015}
{Arregui}, I. 2015, Philosophical Transactions of the Royal Society of London
  Series A, 373, 20140261, \dodoi{10.1098/rsta.2014.0261}

\bibitem[{{Aschwanden} {et~al.}(1999){Aschwanden}, {Fletcher}, {Schrijver}, \&
  {Alexander}}]{aschwanden1999}
{Aschwanden}, M.~J., {Fletcher}, L., {Schrijver}, C.~J., \& {Alexander}, D.
  1999, \apj, 520, 880.
\newblock
  \url{http://adsabs.harvard.edu/cgi-bin/nph-bib_query?bibcode=1999ApJ...520..880A&db_key=AST}

\bibitem[{De~Pontieu {et~al.}(2007)De~Pontieu, McIntosh, Carlsson, Hansteen,
  Tarbell, Schrijver, Title, Shine, Tsuneta, Katsukawa, Ichimoto, Suematsu,
  Shimizu, \& Nagata}]{DePontieu2007}
De~Pontieu, B., McIntosh, S., Carlsson, M., {et~al.} 2007, Science (New York,
  N.Y.), 318, 1574—1577, \dodoi{10.1126/science.1151747}

\bibitem[{{Fontenla} {et~al.}(2007){Fontenla}, {Balasubramaniam}, \&
  {Harder}}]{2007ApJ...667.1243F}
{Fontenla}, J.~M., {Balasubramaniam}, K.~S., \& {Harder}, J. 2007, \apj, 667,
  1243, \dodoi{10.1086/520319}

\bibitem[{{Garc{\'\i}a Marirrodriga} {et~al.}(2021){Garc{\'\i}a Marirrodriga},
  {Pacros}, {Strandmoe}, {Arcioni}, {Arts}, {Ashcroft}, {Ayache}, {Bonnefous},
  {Brahimi}, {Cipriani}, {Damasio}, {De Jong}, {D{\'e}prez}, {Fahmy}, {Fels},
  {Fiebrich}, {Hass}, {Hern{\'a}ndez}, {Icardi}, {Junge}, {Kletzkine}, {Laget},
  {Le Deuff}, {Liebold}, {Lodiot}, {Marliani}, {Mascarello}, {M{\"u}ller},
  {Oganessian}, {Olivier}, {Palombo}, {Philippe}, {Ragnit}, {Ramachandran},
  {S{\'a}nchez P{\'e}rez}, {Stienstra}, {Th{\"u}rey}, {Urwin}, {Wirth}, \&
  {Zouganelis}}]{Marirrodriga2021}
{Garc{\'\i}a Marirrodriga}, C., {Pacros}, A., {Strandmoe}, S., {et~al.} 2021,
  \aap, 646, A121, \dodoi{10.1051/0004-6361/202038519}

\bibitem[{{Goossens} {et~al.}(2012){Goossens}, {Andries}, {Soler}, {Van
  Doorsselaere}, {Arregui}, \& {Terradas}}]{Goossens2012}
{Goossens}, M., {Andries}, J., {Soler}, R., {et~al.} 2012, \apj, 753, 111,
  \dodoi{10.1088/0004-637X/753/2/111}

\bibitem[{{Goossens} {et~al.}(2013){Goossens}, {Van Doorsselaere}, {Soler}, \&
  {Verth}}]{Goossens2013}
{Goossens}, M., {Van Doorsselaere}, T., {Soler}, R., \& {Verth}, G. 2013, \apj,
  771, 74, \dodoi{10.1088/0004-637X/771/1/74}

\bibitem[{{Hillier} {et~al.}(2020){Hillier}, {Van Doorsselaere}, \&
  {Karampelas}}]{Hillier2020}
{Hillier}, A., {Van Doorsselaere}, T., \& {Karampelas}, K. 2020, \apjl, 897,
  L13, \dodoi{10.3847/2041-8213/ab9ca3}

\bibitem[{{Jess} {et~al.}(2009){Jess}, {Mathioudakis}, {Erd{\'e}lyi},
  {Crockett}, {Keenan}, \& {Christian}}]{Jess2009}
{Jess}, D.~B., {Mathioudakis}, M., {Erd{\'e}lyi}, R., {et~al.} 2009, Science,
  323, 1582, \dodoi{10.1126/science.1168680}

\bibitem[{{Karampelas} {et~al.}(2017){Karampelas}, {Van Doorsselaere}, \&
  {Antolin}}]{karampelas2017}
{Karampelas}, K., {Van Doorsselaere}, T., \& {Antolin}, P. 2017, \aap, 604,
  A130, \dodoi{10.1051/0004-6361/201730598}

\bibitem[{{Karampelas} {et~al.}(2019){Karampelas}, {Van Doorsselaere},
  {Pascoe}, {Guo}, \& {Antolin}}]{Karampelas2019}
{Karampelas}, K., {Van Doorsselaere}, T., {Pascoe}, D.~J., {Guo}, M., \&
  {Antolin}, P. 2019, Frontiers in Astronomy and Space Sciences, 6, 38,
  \dodoi{10.3389/fspas.2019.00038}

\bibitem[{{Keppens} {et~al.}(2020){Keppens}, {Teunissen}, {Xia}, \&
  {Porth}}]{2020arXiv200403275K}
{Keppens}, R., {Teunissen}, J., {Xia}, C., \& {Porth}, O. 2020, arXiv e-prints,
  arXiv:2004.03275.
\newblock \doarXiv{2004.03275}

\bibitem[{{Kobayashi} {et~al.}(2014){Kobayashi}, {Cirtain}, {Winebarger},
  {Korreck}, {Golub}, {Walsh}, {De Pontieu}, {DeForest}, {Title}, {Kuzin},
  {Savage}, {Beabout}, {Beabout}, {Podgorski}, {Caldwell}, {McCracken},
  {Ordway}, {Bergner}, {Gates}, {McKillop}, {Cheimets}, {Platt}, {Mitchell}, \&
  {Windt}}]{Kobayashi2014}
{Kobayashi}, K., {Cirtain}, J., {Winebarger}, A.~R., {et~al.} 2014, \solphys,
  289, 4393, \dodoi{10.1007/s11207-014-0544-4}

\bibitem[{{Kolotkov} {et~al.}(2018){Kolotkov}, {Nakariakov}, \&
  {Kontar}}]{Kolotkov2018}
{Kolotkov}, D.~Y., {Nakariakov}, V.~M., \& {Kontar}, E.~P. 2018, \apj, 861, 33,
  \dodoi{10.3847/1538-4357/aac77e}

\bibitem[{{Korendyke} {et~al.}(2001){Korendyke}, {Vourlidas}, {Cook}, {Dere},
  {Howard}, {Morrill}, {Moses}, {Moulton}, \& {Socker}}]{Korendyke2001}
{Korendyke}, C.~M., {Vourlidas}, A., {Cook}, J.~W., {et~al.} 2001, \solphys,
  200, 63, \dodoi{10.1023/A:1010310217570}

\bibitem[{{Magyar} \& {Nakariakov}(2020)}]{2020ApJ...894L..23M}
{Magyar}, N., \& {Nakariakov}, V.~M. 2020, \apjl, 894, L23,
  \dodoi{10.3847/2041-8213/ab8e36}

\bibitem[{{McIntosh} {et~al.}(2011){McIntosh}, {de Pontieu}, {Carlsson},
  {Hansteen}, {Boerner}, \& {Goossens}}]{McIntosh2011}
{McIntosh}, S.~W., {de Pontieu}, B., {Carlsson}, M., {et~al.} 2011, \nat, 475,
  477, \dodoi{10.1038/nature10235}

\bibitem[{{Morton} {et~al.}(2016){Morton}, {Tomczyk}, \& {Pinto}}]{Morton2016}
{Morton}, R.~J., {Tomczyk}, S., \& {Pinto}, R.~F. 2016, \apj, 828, 89,
  \dodoi{10.3847/0004-637X/828/2/89}

\bibitem[{{Morton} {et~al.}(2019){Morton}, {Weberg}, \&
  {McLaughlin}}]{Morton2019}
{Morton}, R.~J., {Weberg}, M.~J., \& {McLaughlin}, J.~A. 2019, Nature
  Astronomy, 3, 223, \dodoi{10.1038/s41550-018-0668-9}

\bibitem[{{Nakariakov} {et~al.}(1999){Nakariakov}, {Ofman}, {DeLuca},
  {Roberts}, \& {Davila}}]{nakariakov1999}
{Nakariakov}, V.~M., {Ofman}, L., {DeLuca}, E.~E., {Roberts}, B., \& {Davila},
  J.~M. 1999, Sci., 285, 862

\bibitem[{{Nistic{\`o}} {et~al.}(2013){Nistic{\`o}}, {Nakariakov}, \&
  {Verwichte}}]{nistico2013}
{Nistic{\`o}}, G., {Nakariakov}, V.~M., \& {Verwichte}, E. 2013, \aap, 552,
  A57, \dodoi{10.1051/0004-6361/201220676}

\bibitem[{{Rochus} {et~al.}(2020){Rochus}, {Auch{\`e}re}, {Berghmans}, {Harra},
  {Schmutz}, {Sch{\"u}hle}, {Addison}, {Appourchaux}, {Aznar Cuadrado},
  {Baker}, {Barbay}, {Bates}, {BenMoussa}, {Bergmann}, {Beurthe}, {Borgo},
  {Bonte}, {Bouzit}, {Bradley}, {B{\"u}chel}, {Buchlin}, {B{\"u}chner},
  {Cab{\'e}}, {Cadiergues}, {Chaigneau}, {Chares}, {Choque Cortez}, {Coker},
  {Condamin}, {Coumar}, {Curdt}, {Cutler}, {Davies}, {Davison}, {Defise}, {Del
  Zanna}, {Delmotte}, {Delouille}, {Dolla}, {Dumesnil}, {D{\"u}rig}, {Enge},
  {Fran{\c{c}}ois}, {Fourmond}, {Gillis}, {Giordanengo}, {Gissot}, {Green},
  {Guerreiro}, {Guilbaud}, {Gyo}, {Haberreiter}, {Hafiz}, {Hailey}, {Halain},
  {Hansotte}, {Hecquet}, {Heerlein}, {Hellin}, {Hemsley}, {Hermans}, {Hervier},
  {Hochedez}, {Houbrechts}, {Ihsan}, {Jacques}, {J{\'e}r{\^o}me}, {Jones},
  {Kahle}, {Kennedy}, {Klaproth}, {Kolleck}, {Koller}, {Kotsialos},
  {Kraaikamp}, {Langer}, {Lawrenson}, {Le Clech'}, {Lenaerts}, {Liebecq},
  {Linder}, {Long}, {Mampaey}, {Markiewicz-Innes}, {Marquet}, {Marsch},
  {Matthews}, {Mazy}, {Mazzoli}, {Meining}, {Meltchakov}, {Mercier}, {Meyer},
  {Monecke}, {Monfort}, {Morinaud}, {Moron}, {Mountney}, {M{\"u}ller},
  {Nicula}, {Parenti}, {Peter}, {Pfiffner}, {Philippon}, {Phillips},
  {Plesseria}, {Pylyser}, {Rabecki}, {Ravet-Krill}, {Rebellato}, {Renotte},
  {Rodriguez}, {Roose}, {Rosin}, {Rossi}, {Roth}, {Rouesnel}, {Roulliay},
  {Rousseau}, {Ruane}, {Scanlan}, {Schlatter}, {Seaton}, {Silliman}, {Smit},
  {Smith}, {Solanki}, {Spescha}, {Spencer}, {Stegen}, {Stockman}, {Szwec},
  {Tamiatto}, {Tandy}, {Teriaca}, {Theobald}, {Tychon}, {van Driel-Gesztelyi},
  {Verbeeck}, {Vial}, {Werner}, {West}, {Westwood}, {Wiegelmann}, {Willis},
  {Winter}, {Zerr}, {Zhang}, \& {Zhukov}}]{Rochus2020}
{Rochus}, P., {Auch{\`e}re}, F., {Berghmans}, D., {et~al.} 2020, \aap, 642, A8,
  \dodoi{10.1051/0004-6361/201936663}

\bibitem[{{Samanta} {et~al.}(2016){Samanta}, {Singh}, {Sindhuja}, \&
  {Banerjee}}]{Samanta2016}
{Samanta}, T., {Singh}, J., {Sindhuja}, G., \& {Banerjee}, D. 2016, \solphys,
  291, 155, \dodoi{10.1007/s11207-015-0821-x}

\bibitem[{{Schrijver} {et~al.}(1999){Schrijver}, {Title}, {Berger}, {Fletcher},
  {Hurlburt}, {Nightingale}, {Shine}, {Tarbell}, {Wolfson}, {Golub},
  {Bookbinder}, {Deluca}, {McMullen}, {Warren}, {Kankelborg}, {Handy}, \& {de
  Pontieu}}]{schrijver1999}
{Schrijver}, C.~J., {Title}, A.~M., {Berger}, T.~E., {et~al.} 1999, \solphys,
  187, 261

\bibitem[{{Shi} {et~al.}(2021){Shi}, {Van Doorsselaere}, {Guo}, {Karampelas},
  {Li}, \& {Antolin}}]{shi2021}
{Shi}, M., {Van Doorsselaere}, T., {Guo}, M., {et~al.} 2021, \apj, 908, 233,
  \dodoi{10.3847/1538-4357/abda54}

\bibitem[{{Terradas} {et~al.}(2008){Terradas}, {Andries}, {Goossens},
  {Arregui}, {Oliver}, \& {Ballester}}]{Terradas2008}
{Terradas}, J., {Andries}, J., {Goossens}, M., {et~al.} 2008, \apjl, 687, L115,
  \dodoi{10.1086/593203}

\bibitem[{{Terradas} \& {Arregui}(2018)}]{terradas2018b}
{Terradas}, J., \& {Arregui}, I. 2018, Research Notes of the American
  Astronomical Society, 2, 196, \dodoi{10.3847/2515-5172/aaeb26}

\bibitem[{{The SunPy Community} {et~al.}(2020){The SunPy Community}, Barnes,
  Bobra, Christe, Freij, Hayes, Ireland, Mumford, Perez-Suarez, Ryan, Shih,
  Chanda, Glogowski, Hewett, Hughitt, Hill, Hiware, Inglis, Kirk, Konge, Mason,
  Maloney, Murray, Panda, Park, Pereira, Reardon, Savage, Sipőcz, Stansby,
  Jain, Taylor, Yadav, Rajul, \& Dang}]{sunpy2020}
{The SunPy Community}, Barnes, W.~T., Bobra, M.~G., {et~al.} 2020, The
  Astrophysical Journal, 890, 68, \dodoi{10.3847/1538-4357/ab4f7a}

\bibitem[{{Thurgood} {et~al.}(2014){Thurgood}, {Morton}, \&
  {McLaughlin}}]{Thurgood2014}
{Thurgood}, J.~O., {Morton}, R.~J., \& {McLaughlin}, J.~A. 2014, \apjl, 790,
  L2, \dodoi{10.1088/2041-8205/790/1/L2}

\bibitem[{{Tian} {et~al.}(2012){Tian}, {McIntosh}, {Wang}, {Ofman}, {De
  Pontieu}, {Innes}, \& {Peter}}]{tian2012}
{Tian}, H., {McIntosh}, S.~W., {Wang}, T., {et~al.} 2012, \apj, 759, 144,
  \dodoi{10.1088/0004-637X/759/2/144}

\bibitem[{Tomczyk {et~al.}(2007)Tomczyk, McIntosh, Keil, Judge, Schad, Seeley,
  \& Edmondson}]{Tomczyk2007}
Tomczyk, S., McIntosh, S.~W., Keil, S.~L., {et~al.} 2007, Science, 317, 1192,
  \dodoi{10.1126/science.1143304}

\bibitem[{{Van Doorsselaere} {et~al.}(2014){Van Doorsselaere}, {Gijsen},
  {Andries}, \& {Verth}}]{Doorsselaere2014}
{Van Doorsselaere}, T., {Gijsen}, S.~E., {Andries}, J., \& {Verth}, G. 2014,
  \apj, 795, 18, \dodoi{10.1088/0004-637X/795/1/18}

\bibitem[{{Van Doorsselaere} {et~al.}(2008){Van Doorsselaere}, {Nakariakov}, \&
  {Verwichte}}]{vd2008}
{Van Doorsselaere}, T., {Nakariakov}, V.~M., \& {Verwichte}, E. 2008, \apjl,
  676, L73

\bibitem[{{Van Doorsselaere} {et~al.}(2020){Van Doorsselaere}, {Srivastava},
  {Antolin}, {Magyar}, {Vasheghani Farahani}, {Tian}, {Kolotkov}, {Ofman},
  {Guo}, {Arregui}, {De Moortel}, \& {Pascoe}}]{Doorsselaere2020}
{Van Doorsselaere}, T., {Srivastava}, A.~K., {Antolin}, P., {et~al.} 2020,
  \ssr, 216, 140, \dodoi{10.1007/s11214-020-00770-y}

\bibitem[{{Wang} {et~al.}(2012){Wang}, {Ofman}, {Davila}, \& {Su}}]{wang2012}
{Wang}, T., {Ofman}, L., {Davila}, J.~M., \& {Su}, Y. 2012, \apjl, 751, L27,
  \dodoi{10.1088/2041-8205/751/2/L27}

\bibitem[{{Williams} {et~al.}(2002){Williams}, {Mathioudakis}, {Gallagher},
  {Phillips}, {McAteer}, {Keenan}, {Rudawy}, \& {Katsiyannis}}]{Williams2002}
{Williams}, D.~R., {Mathioudakis}, M., {Gallagher}, P.~T., {et~al.} 2002,
  \mnras, 336, 747, \dodoi{10.1046/j.1365-8711.2002.05764.x}

\bibitem[{{Withbroe} \& {Noyes}(1977)}]{Withbroe1977}
{Withbroe}, G.~L., \& {Noyes}, R.~W. 1977, \araa, 15, 363,
  \dodoi{10.1146/annurev.aa.15.090177.002051}

\bibitem[{{Xia} {et~al.}(2018){Xia}, {Teunissen}, {El Mellah}, {Chan{\'e}}, \&
  {Keppens}}]{2018ApJS..234...30X}
{Xia}, C., {Teunissen}, J., {El Mellah}, I., {Chan{\'e}}, E., \& {Keppens}, R.
  2018, \apjs, 234, 30, \dodoi{10.3847/1538-4365/aaa6c8}

\end{thebibliography}
\bibliographystyle{aasjournal}



\end{document}